\documentclass[
 reprint,
 amsmath,amssymb,
 aps,
 physrev,
]{revtex4-2}

\usepackage{color}
\usepackage{graphicx}
\usepackage{dcolumn}
\usepackage{bm}
\usepackage{braket}
\usepackage{empheq}
\usepackage{natbib}
\usepackage{makecell}
\usepackage[nopatch=footnote]{microtype}

\begin{document}

\title{Sequential optimization for orbital-optimized variational quantum eigensolvers}

\author{Hiromichi Hayashi}
\email{hayashi-hiromichi878@g.ecc.u-tokyo.ac.jp}

\author{Takeshi Sato}
\email[Contact author: ]{sato@atto.t.u-tokyo.ac.jp}
\affiliation{%
    Graduate School of Engineering, The University of Tokyo, Tokyo, Japan
}%

\date{\today}

\begin{abstract}
Orbital optimization within the variational quantum eigensolver (VQE) framework is useful for improving active-space quality and accurately capturing electron correlation with shallow quantum circuits. The orbital-optimized unitary coupled cluster (OO-UCC) method [W. Mizukami \textit{et al.}, Phys. Rev. Res. \textbf{2}, 033421 (2020)], in its original formulation, employs the orbital gradient and Hessian in a Newton–Raphson procedure. In this study, we propose sequential orbital-optimized unitary coupled cluster doubles (SOO-UCCD), extending the gradient-free sequential minimal optimization approach [K. M. Nakanishi \textit{et al.}, Phys. Rev. Res. \textbf{2}, 043158 (2020)] to orbital rotation parameters. Based on the frequency structure implied by the algebraic properties of the orbital rotation generators, our method determines each orbital rotation angle from a nine-point energy profile. The energy landscape is evaluated by classically transforming the Hamiltonian integrals and contracting them with the reduced density matrices already measured for the VQE energy evaluation. This allows for orbital optimization without evaluating the orbital gradient or Hessian. By building on the core benefits of the OO-UCC framework, our method provides a gradient-free and hyperparameter-free orbital-optimization procedure. State-vector simulations for LiH in the STO-3G basis show that SOO-UCCD reproduces UCCSD-level accuracy along the bond-stretching curve and exhibits macro-iteration convergence comparable to that of L-BFGS-based orbital optimization.
\end{abstract}

\maketitle

\section{Introduction}
The variational quantum eigensolver (VQE) provides a quantum--classical variational framework for molecular electronic-structure calculations~\cite{peruzzo2014,mcclean2016,omalley2016}.
It combines a parameterized quantum circuit for preparing a trial wavefunction with a classical optimization loop that minimizes the measured energy.
Efficient wavefunction parameterizations and optimization procedures remain important for reducing quantum computational resources not only on near-term devices but also in fault-tolerant architectures~\cite{sayginel2024}.
Among the various wavefunction ans\"atze proposed for molecular VQE, the unitary coupled cluster (UCC) approach has been widely studied as a physically motivated construction based on fermionic excitation operators.
Experimental demonstrations and numerical simulations have shown the potential of UCC-based VQE for molecular energy calculations~\cite{omalley2016,romero2019}.

In conventional UCC-based VQE calculations, however, the molecular orbitals defining the reference determinant and excitation operators are typically fixed and are often taken from a Hartree--Fock calculation. Although the cluster amplitudes are determined variationally, the resulting wavefunction is not fully variational with respect to orbital rotations. Hartree--Fock orbitals are not generally optimal for a correlated wavefunction, particularly when orbital-relaxation effects are important. This limitation is especially relevant in active-space quantum chemistry calculations, where the choice and optimization of the active orbitals influence the balance between wavefunction accuracy and quantum-resource requirements.
Orbital-optimized UCC (OO-UCC) methods have consequently been proposed within the VQE framework, in which both the cluster amplitudes and molecular orbitals are variationally optimized~\cite{mizukami2020}.
Orbital optimization can improve the quality of a restricted active space, allowing a target accuracy to be reached with fewer qubits and shallower quantum circuits~\cite{takeshita2020,mizukami2020}.

A commonly used approach to orbital optimization in multiconfigurational electronic-structure methods~\cite{siegbahn1980,siegbahn1981,szalay2012,fossotande2016}, and the approach adopted in the OO-UCC formulation~\cite{mizukami2020}, is a Newton--Raphson procedure based on the orbital energy gradient and Hessian. Newton--Raphson methods exhibit quadratic convergence sufficiently close to a solution, but rely on a local second-order approximation to the energy surface.
When the current orbitals are far from the solution, small Hessian eigenvalues can produce excessively large orbital updates, while an indefinite Hessian can invalidate the local quadratic minimization model. Consequently, practical multiconfigurational self-consistent-field implementations may require Hessian modifications, step-length restrictions, mode damping, or related stabilization procedures~\cite{werner1980,siegbahn1981,yeager1982}.
In quantum-computing implementations, finite-sampling statistical errors in the measured reduced density matrices can also affect the orbital gradients and Hessians.

An alternative strategy is to optimize the orbitals through successive two-orbital rotations. In classical multiconfigurational self-consistent-field theory, orbital-optimization methods based on sequential Jacobi rotations have been developed in which the energy is minimized with respect to one rotation angle at a time~\cite{ivanic2003}.
Unlike a conventional Newton--Raphson step, this approach does not rely on a multidimensional local quadratic approximation. Instead, the exact one-dimensional energy dependence associated with each orbital pair is exploited directly.

A closely related concept has recently been used for the optimization of parameterized quantum circuits. Sequential minimal optimization~\cite{nakanishi2020} reconstructs the analytic energy dependence on an individual circuit parameter and updates that parameter to the minimum of the reconstructed function.
More recently, ExcitationSolve~\cite{jager2025} extended this idea to fermionic excitation operators, whose generators produce a second-order Fourier energy landscape along each parameter direction. It also supports Givens rotations and shared/repeated parameters, for which higher-order finite Fourier representations arise.
ExcitationSolve is gradient-free and hyperparameter-free and uses global information along the selected coordinate rather than only the local slope.

In this work, we develop sequential orbital-optimized unitary coupled cluster doubles (SOO-UCCD), applying gradient-free sequential energy-profile reconstruction to both quantum-circuit and orbital-rotation parameters within the VQE framework. For each spin-restricted spatial orbital rotation, we formulate a finite Fourier representation and determine its minimum from nine energy evaluations. These energies are obtained classically by transforming the one- and two-electron integrals and contracting them with the reduced density matrices measured for the current VQE state. The resulting orbital-optimization procedure avoids explicit orbital-gradient and Hessian evaluations while retaining the coordinate-wise globally informed and hyperparameter-free character of sequential minimal optimization.

The remainder of this paper is organized as follows. Section~\ref{sec:theory} presents the UCC ansatz and orbital rotations, derives the finite Fourier representations of the one-parameter energy profiles, and describes their minimization and the optimization workflow. Section~\ref{sec:results} evaluates the accuracy and convergence of SOO-UCCD through state-vector simulations of LiH. Section~\ref{sec:conclusion} summarizes the findings.

\section{Theory}
\label{sec:theory}
\subsection{Unitary Coupled Cluster Method}
The Hamiltonian of an electronic system in second quantization is generally given by:
\begin{equation}
    \label{eq:hamiltonian}
    \hat{H} = \sum_{pq} h_{pq} \hat{a}_p^\dagger \hat{a}_q + \frac{1}{2} \sum_{pqrs} h_{pqrs} \hat{a}_p^\dagger \hat{a}_q^\dagger \hat{a}_s \hat{a}_r,
\end{equation}
where \( h_{pq} \) and \( h_{pqrs} \) are the one- and two-electron integrals in the molecular orbital (MO) basis, respectively.
In this work, we consider spin-restricted closed-shell systems, where \( \alpha \) and \( \beta \) spin components share the same spatial molecular orbitals.
In the UCC framework~\cite{romero2019}, the wavefunction is parameterized as:
\begin{equation}
    \label{eq:ucc_wavefunction}
    |\Psi(\theta)\rangle = e^{\hat{A}(\theta)} |\Phi_0\rangle,
\end{equation}
where \( |\Phi_0\rangle  \) is a reference state (typically the Hartree--Fock state) and \( \hat{A}(\theta) \) is an anti-Hermitian operator defined as \( \hat{A} = \hat{T} - \hat{T}^\dagger  \) using the cluster operator \( \hat{T} \).
The cluster operator typically consists of single (\( \hat{T}_1 \)) and double (\( \hat{T}_2 \)) excitations, which are defined as:
\begin{equation}
    \label{eq:T1}
    \hat{T}_1 = \sum_{\substack{i \in \mathrm{occ},\, a \in \mathrm{virt} \\ \Delta S_z = 0}}
    \theta_i^a \hat{a}_a^\dagger \hat{a}_i,
\end{equation}
\begin{equation}
    \label{eq:T2}
    \hat{T}_2 = \sum_{\substack{i < j,\, a < b \\ \Delta S_z = 0}}
    \theta_{ij}^{ab} \hat{a}_a^\dagger \hat{a}_b^\dagger \hat{a}_j \hat{a}_i,
\end{equation}
In these expressions, \(i,j\) denote occupied and \(a,b\) virtual spin-orbital indices in the reference state \(|\Phi_0\rangle\).
\color{black}
The spin orbitals are initially ordered by increasing Hartree--Fock orbital energy, with each spatial orbital represented by consecutive \(\alpha\)- and \(\beta\)-spin indices; thus, even (odd) indices correspond to \(\alpha\) (\(\beta\)) spin. The spin-orbital indices are kept fixed throughout the orbital-optimization procedure.
\color{black}
The constraint \( \Delta S_z = 0 \) ensures conservation of total spin projection, and the ordered summation \( i < j \), \( a < b \) in \( \hat{T}_2 \) eliminates redundant degrees of freedom arising from fermionic antisymmetry. Thus, the number of independent variational parameters exactly matches the number of unique excitation pairs implemented on the quantum circuit.

In this study, we use the first-order product (disentangled) form of the UCC ansatz,
\begin{equation}
    \label{eq:product_ucc}
    |\Psi(\bm{\theta})\rangle
    =
    \prod_{\mu}
    \exp\!\left[\theta_{\mu}\left(\hat{\tau}_{\mu}-\hat{\tau}_{\mu}^{\dagger}\right)\right]
    |\Phi_0\rangle,
\end{equation}
where \( \theta_{\mu} \) and \( \hat{\tau}_{\mu} \) denote, respectively, an individual amplitude and excitation operator appearing in \( \hat{T}_1 \) or \( \hat{T}_2 \).
\color{black}
The excitation operators in the product are applied in a fixed order, with all single excitations applied first, followed by all double excitations. Within each excitation rank, the excitation operators are ordered lexicographically by the orbital indicies: \((i, a)\) for singles and \((i, j, a, b)\) for doubles, with \(i < j\) and \(a < b\).
\color{black}
We primarily employ the product-form unitary coupled cluster doubles (UCCD) ansatz as the base of our proposed method, retaining only double-excitation gates. For comparison and to assess the effect of orbital optimization, we also consider the corresponding product-form unitary coupled cluster singles and doubles (UCCSD) ansatz.
\subsection{Orbital Optimization via Sequential Givens Rotations}
Orbital optimization~\cite{ivanic2003,mizukami2020} is introduced as a unitary transformation in the one-particle orbital space. Let \( \boldsymbol{\kappa} \) be an anti-Hermitian orbital-rotation matrix satisfying \( \boldsymbol{\kappa}^{\dagger}=-\boldsymbol{\kappa} \). The rotated spatial molecular orbitals are defined as
\begin{equation}
    \label{eq:oo_unitary_rotation}
    \tilde{\phi}_p
    =
    \sum_q
    \phi_q U_{qp},
    \qquad
    \mathbf{U}
    :=
    e^{\boldsymbol{\kappa}} .
\end{equation}
The corresponding transformation in Fock space is generated by the one-body anti-Hermitian operator
\begin{equation}
    \hat{\kappa}
    =
    \sum_{pq}
    \kappa_{pq}
    \hat{a}_{p}^{\dagger}
    \hat{a}_{q}.
\end{equation}
With this correspondence, the orbital rotation can be implemented either as a transformation of the molecular spin orbitals or, equivalently, as a similarity transformation of operators in Fock space.
The transformation of the creation and annihilation operators is given by:
\begin{equation}
    \tilde{a}_{p}^\dagger
    :=
    e^{\hat{\kappa}}
    \hat{a}_{p}^\dagger
    e^{-\hat{\kappa}}
    =
    \sum_q
    \hat{a}_{q}^\dagger U_{qp},
\end{equation}
\begin{equation}
    \tilde{a}_{p}
    :=
    e^{\hat{\kappa}}
    \hat{a}_{p}
    e^{-\hat{\kappa}}
    =
    \sum_q
    \hat{a}_{q} U_{qp}^*.
\end{equation}
By substituting these operators, the one-electron term of the Hamiltonian transforms as:
\begin{align}
    \label{eq:one_electron_transform}
    e^{\hat{\kappa}} \left( \sum_{pq} h_{pq} \hat{a}_p^\dagger \hat{a}_q \right) e^{-\hat{\kappa}}
     & = \sum_{pq} h_{pq} \tilde{a}_p^\dagger \tilde{a}_q \nonumber         \\
     & = \sum_{tu} \tilde{h}_{tu} \hat{a}_{t}^\dagger \hat{a}_{u} \nonumber \\
     & = \sum_{pq} \tilde{h}_{pq} \hat{a}_p^\dagger \hat{a}_q,
\end{align}
where the rotated one-electron integrals are \( \tilde{h}_{pq} = \sum_{tu} U_{pt} h_{tu} U_{qu}^*  \). Similarly, the two-electron term transforms as:
\begin{equation}
    \label{eq:two_electron_transform}
    \begin{split}
        e^{\hat{\kappa}} \left( \frac{1}{2} \sum_{pqrs} h_{pqrs} \hat{a}_p^\dagger \hat{a}_q^\dagger \hat{a}_s \hat{a}_r \right) e^{-\hat{\kappa}}
         & = \frac{1}{2} \sum_{pqrs} h_{pqrs} \tilde{a}_p^\dagger \tilde{a}_q^\dagger \tilde{a}_s \tilde{a}_r         \\
         & = \frac{1}{2} \sum_{tuvw} \tilde{h}_{tuvw} \hat{a}_{t}^\dagger \hat{a}_{u}^\dagger \hat{a}_{w} \hat{a}_{v} \\
         & = \frac{1}{2} \sum_{pqrs} \tilde{h}_{pqrs} \hat{a}_p^\dagger \hat{a}_q^\dagger \hat{a}_s \hat{a}_r,
    \end{split}
\end{equation}
where the rotated two-electron integrals are explicitly defined as \( \tilde{h}_{pqrs} = \sum_{tuvw} U_{pt} U_{qu} h_{tuvw} U_{rv}^* U_{sw}^* \).
Consequently, the total energy is expressed as:
\begin{equation}
    \label{eq:energy_kappa}
    \begin{split}
        E(\bm{\theta}, \bm{\kappa}) & = \langle \Psi(\bm{\theta}) | \, e^{\hat{\kappa}} \hat{H} e^{-\hat{\kappa}} \, | \Psi(\bm{\theta}) \rangle \\
                                    & = \langle \Psi(\bm{\theta}) | \, \bar{H}(\bm{\kappa}) \, | \Psi(\bm{\theta}) \rangle,
    \end{split}
\end{equation}
where \( \bar{H}(\bm{\kappa}) \) is the transformed Hamiltonian determined by the rotated one-electron integrals \( \tilde{h}_{pq}(\bm{\kappa}) \) and two-electron integrals \( \tilde{h}_{pqrs}(\bm{\kappa}) \).
For the sequential optimization, we use spin-restricted occupied-virtual orbital-rotation parameters. The corresponding anti-Hermitian operator is defined as
\begin{equation}
    \label{eq:spin_summed_kappa}
    \hat{\kappa}
    =
    \sum_{ia}
    \kappa_{ia}
    \sum_{\sigma}
    \left(
    \hat{a}_{a\sigma}^{\dagger}
    \hat{a}_{i\sigma}
    -
    \hat{a}_{i\sigma}^{\dagger}
    \hat{a}_{a\sigma}
    \right),
\end{equation}
where \( \kappa_{ia} \) is a real parameter, and \( i \) and \( a \) denote occupied and virtual spatial orbital indices, respectively.

To circumvent the computational hurdle of a monolithic gradient-based optimization over \( \bm{\kappa} \), we implement a pairwise sequential sweep algorithm. Each parameter \( \kappa_{ia} \) is updated sequentially, while the effects of previously optimized parameters are absorbed into the transformed electron integrals. Therefore, at each update step, it is sufficient to consider a generator in which only one spatial orbital pair \( (a, i) \) has a nonzero parameter.
For a single-pair update, the orbital rotation is a one-parameter Givens rotation that mixes only the occupied orbital \( i \) and the virtual orbital \( a  \). This sparsity allows the electron integrals to be updated in place, with a cost of \( \mathcal{O}(N) \) for the one-electron integrals and \( \mathcal{O}(N^3) \) for the two-electron integrals, where \( N \) is the number of spatial orbitals.
\subsection{Frequency Analysis of Fermionic Generators}
\subsubsection{Algebraic Properties of a Fermionic Excitation Generator}
The orbital-rotation generators in Eq.~\eqref{eq:spin_summed_kappa} and fermionic excitation operators in Eq.~\eqref{eq:T1} and Eq.~\eqref{eq:T2} share a common algebraic structure.
Let \( \mu \) label a generator component and define the rank-\( k \) excitation operator
\begin{equation}
    \hat{\tau}_{\mu}
    =
    \hat{a}_{p_{\mu 1}}^{\dagger}
    \cdots
    \hat{a}_{p_{\mu k}}^{\dagger}
    \hat{a}_{q_{\mu k}}
    \cdots
    \hat{a}_{q_{\mu 1}},
\end{equation}
where \( P_{\mu}=\{p_{\mu 1},\ldots,p_{\mu k}\} \) and \( Q_{\mu}=\{q_{\mu 1},\ldots,q_{\mu k}\} \) are ordered tuples of distinct fermionic-mode indices, with no index appearing in both \( P_\mu \) and \( Q_\mu \).
The corresponding Hermitian generator is
\begin{equation}
    \label{eq:generator}
    \hat{G}_{\mu} := i\left(\hat{\tau}_{\mu}-\hat{\tau}_{\mu}^{\dagger}\right).
\end{equation}
The operators \( \hat{\tau}_\mu \) and \( \hat{\tau}_\mu^\dagger \) connect two disjoint occupation subspaces, characterized by opposite occupations on \( P_\mu \) and \( Q_\mu \), and annihilate all other determinants. Here
\begin{equation}
    \begin{aligned}
        \hat{\Pi}_{\mu}
        =\hat{G}_{\mu}^{2}
         & =\hat{\tau}_{\mu}\hat{\tau}_{\mu}^{\dagger}
        +\hat{\tau}_{\mu}^{\dagger}\hat{\tau}_{\mu}    \\
         & =\prod_{p\in P_{\mu}}\hat{n}_{p}
        \prod_{q\in Q_{\mu}}(1-\hat{n}_{q})            \\
         & \quad+\prod_{q\in Q_{\mu}}\hat{n}_{q}
        \prod_{p\in P_{\mu}}(1-\hat{n}_{p})
    \end{aligned}
\end{equation}
is the projection operator onto these two subspaces.
Thus, \( \hat{\Pi}_{\mu} \) satisfies
\begin{equation}
    \hat{\Pi}_{\mu}^{2}=\hat{\Pi}_{\mu},
    \qquad
    \hat{G}_{\mu}\hat{\Pi}_{\mu}
    =
    \hat{\Pi}_{\mu}\hat{G}_{\mu}
    =
    \hat{G}_{\mu}.
\end{equation}
It follows that the generalized excitation generator satisfies the cubic relation
\begin{equation}
    \hat{G}_{\mu}^{3} = \hat{G}_{\mu}.
\end{equation}
The polynomial \( \lambda^3 - \lambda = 0 \) therefore restricts the eigenvalues of \( \hat{G}_{\mu} \) to \( \{0,\pm 1\} \), independently of the excitation rank \( k \).
\subsubsection{Mutually Commuting Generators}
Consider a total generator composed of \( m \) mutually commuting fermionic excitation generators,
\begin{equation}
    \hat{G} = \sum_{r=1}^{m} \hat{G}_{\mu_r},
    \qquad
    [\hat{G}_{\mu_r}, \hat{G}_{\mu_s}] = 0
    \quad (r,s = 1,\dots,m).
\end{equation}
Because the component generators can be simultaneously diagonalized, the eigenvalues \( \lambda \) of \( \hat{G} \) are sums of their individual eigenvalues and thus belong to
\begin{equation}
    \lambda \in \{-m, -(m-1), \dots, 0, \dots, (m-1), m\}.
\end{equation}
Consequently, \( \hat{G} \) satisfies the following operator identity:
\begin{equation}
    \hat{G} \prod_{\ell=1}^m (\hat{G}^2 - \ell^2) = 0.
\end{equation}
By defining \( \{|\lambda\rangle \}  \) as the eigenstates of \( \hat{G} \) with eigenvalues \( \lambda \), and \( \{|\lambda'\rangle \} \) as a basis of spectator states, we can form a complete basis using the tensor product states \( \{|\lambda, \lambda'\rangle \} \). In this representation, \( \hat{G} \) acts solely on the active subspace such that \( \hat{G} |\lambda, \lambda'\rangle = \lambda |\lambda, \lambda'\rangle \).
\subsubsection{Sampling Requirements for One-Parameter Energy Profiles}
The identity operator can be resolved as \( \hat{I} = \sum_{\lambda, \lambda'} |\lambda, \lambda'\rangle \langle \lambda, \lambda'| \). A one-parameter unitary generated by the Hermitian operator \( \hat{G} \) is written as \( U(\varphi) = \exp(-i \varphi \hat{G}) \).
Inserting the identity operators around the Hamiltonian, the energy expectation value is expanded as:
\begin{equation}
    \begin{split}
        E(\varphi) & = \langle \Psi | U^{\dagger}(\varphi) \hat{H} U(\varphi) | \Psi \rangle                                                                                                                                                  \\
                   & = \langle \Psi | e^{i \varphi \hat{G}} \hat{H} e^{-i \varphi \hat{G}} | \Psi \rangle                                                                                                                                     \\
                   & = \sum_{\lambda, \lambda', \eta, \eta'} \langle \Psi | e^{i \varphi \hat{G}} |\lambda, \lambda'\rangle \langle \lambda, \lambda'| \hat{H} |\eta, \eta'\rangle \langle \eta, \eta'| e^{-i \varphi \hat{G}} | \Psi \rangle \\
                   & = \sum_{\lambda, \lambda', \eta, \eta'} e^{i \varphi (\lambda - \eta)} \langle \Psi |\lambda, \lambda'\rangle \langle \lambda, \lambda'| \hat{H} |\eta, \eta'\rangle \langle \eta, \eta'| \Psi \rangle.
    \end{split}
\end{equation}
The possible frequencies \( n = \lambda - \eta \) are integers bounded by \( |n|\leq 2m \). Since the Hamiltonian is Hermitian, the expression can be written as a real finite Fourier series. Thus, \( E(\varphi) \) is a trigonometric polynomial of degree at most \( 2m \):
\begin{equation}
    \label{eq:fourier_expansion}
    E(\varphi) = A_0 + \sum_{n=1}^{2m} \left( A_n \cos(n\varphi) + B_n \sin(n \varphi) \right).
\end{equation}
This expansion contains at most \( 1 + 2(2m) = 4m+1 \) real Fourier coefficients. Therefore, \( 4m+1 \) distinct sampling points are sufficient to determine the one-dimensional energy landscape exactly.

For the spin-restricted OO transformation considered in this work, the orbital-rotation generator is
\begin{equation}
    \label{eq:oo_generator}
    \hat{G}_{\mathrm{OO}}
    =
    \sum_{\sigma \in \{\alpha,\beta \}}
    i
    \left(
    \hat{a}_{a\sigma}^{\dagger}
    \hat{a}_{i\sigma}
    -
    \hat{a}_{i\sigma}^{\dagger}
    \hat{a}_{a\sigma}
    \right),
\end{equation}
which is the sum of two commuting rank-one generators and therefore corresponds to \( m=2 \), requiring \( 4 \times 2 + 1 = 9 \) sampling points.
For the UCC cases, each variational parameter associated with a rank-\(k\) excitation corresponds to a single fermionic excitation generator. This includes the single and double excitations in \(\hat{T}_1\) and \(\hat{T}_2\), defined in Eqs.~\eqref{eq:T1} and~\eqref{eq:T2}, for which \(k=1\) and \(k=2\), respectively. Since each one-parameter update involves only one such generator, independently of the excitation rank \(k\), these cases correspond to \(m=1\) and require \(4\times 1+1=5\) sampling points to determine the corresponding one-dimensional energy profile.

The same algebraic argument also applies to shared-parameter constructions, consistent with the multiple-occurrence generalization of ExcitationSolve~\cite{jager2025}. In particular, consider spin-adapted parameterizations in which two spin-complement excitation generators are tied to a common variational amplitude. For example, one may simultaneously treat the two same-spin double excitations that promote electrons from \(i_\alpha,j_\alpha\) to \(a_\alpha,b_\alpha\) and from \(i_\beta,j_\beta\) to \(a_\beta,b_\beta\). Similarly, one may treat the pair of opposite-spin double excitations that promote electrons from \(i_\alpha,j_\beta\) to \(a_\alpha,b_\beta\) and from \(i_\beta,j_\alpha\) to \(a_\beta,b_\alpha\). In both cases, the paired generators are mutually commuting and form a combined generator with \(m=2\). The corresponding one-dimensional energy profile can therefore be reconstructed from \(4\times 2+1=9\) sampling points, rather than optimizing the two generators separately with \(5+5=10\) samples. More importantly, the simultaneous optimization has the additional advantage of preserving the spin-adapted structure of the ansatz throughout the parameter update, thereby avoiding spin-symmetry breaking that could arise from updating the two spin components independently.
\subsection{Algebraic Minimization via Aberth's Method}
\label{sec:aberth}
This subsection outlines the algebraic method used to find the global minimum of the reconstructed energy function \( E_{ia}(\kappa_{ia}) \) along a single parameter axis.
Since the spin-restricted OO generator in Eq.~\eqref{eq:oo_generator} is the sum of two commuting generators, Eq.~\eqref{eq:fourier_expansion} gives the one-dimensional energy profile as a trigonometric polynomial of degree four:
\begin{equation}
    \label{eq:oo_fourier_expansion}
    E_{ia}(\kappa_{ia}) = A_0 + \sum_{n=1}^{4} \left( {A}_n \cos(n\kappa_{ia}) + {B}_n \sin(n\kappa_{ia}) \right).
\end{equation}
To find the global minimum, we first identify the stationary points satisfying \( E'_{ia}(\kappa_{ia}) = 0 \). Differentiating the above expression yields
\begin{equation}
    E'_{ia}(\kappa_{ia}) = \sum_{n=1}^{4} \left( \tilde{A}_n \cos(n\kappa_{ia}) + \tilde{B}_n \sin(n\kappa_{ia}) \right) = 0,
\end{equation}
where \( \tilde{A}_n = n B_n \) and \( \tilde{B}_n = -n A_n \).
Using Euler's formula and the substitution \( z = e^{i\kappa_{ia}} \), we transform the trigonometric equation into an algebraic equation on the complex plane:
\begin{equation}
    \cos(n\kappa_{ia}) = \frac{z^n + z^{-n}}{2}, \quad \sin(n\kappa_{ia}) = \frac{z^n - z^{-n}}{2i}.
\end{equation}
Substituting these into \( E'_{ia}(\kappa_{ia}) = 0 \) and multiplying both sides by \( z^4 \) yields an associated polynomial \( h(z) \) of degree \( 8 \):
\begin{equation}
    \label{eq:polynomial}
    h(z) = \sum_{n=0}^{8} h_n z^n = 0.
\end{equation}
The roots \( z_j \) of this polynomial that lie on the unit circle (\( |z_j|=1 \)) correspond to the real-valued solutions \( \kappa_{ia}^{(j)} = \arg(z_j) \).
We compute the complex roots of \( h(z)=0 \) using Aberth's method, as implemented in the Rust crate \texttt{aberth}~\cite{aberth}.
By evaluating \( E_{ia}(\kappa_{ia}^{(j)}) \) at these candidates, we identify the global minimum of the reconstructed one-dimensional energy profile without relying on local gradients.
\subsection{Optimization Workflow}
The SOO-UCCD algorithm employs a macro-iteration scheme that alternately optimizes the quantum ansatz amplitudes \( \bm{\theta} \) and the classical orbital parameters \( \bm{\kappa} \).
At the beginning of each macro-iteration, the one- and two-particle reduced density matrices (RDMs), \( \gamma_{pq} \) and \( \Gamma_{pqrs} \), are evaluated on the quantum backend using the current state \( |\Psi(\bm{\theta})\rangle \). We use the spin-summed definitions
\begin{equation}
    \gamma_{pq} = \sum_{\sigma} \langle \Psi(\bm{\theta}) | \hat{a}_{p\sigma}^{\dagger} \hat{a}_{q\sigma} | \Psi(\bm{\theta}) \rangle,
\end{equation}
and
\begin{equation}
    \Gamma_{pqrs} = \sum_{\sigma\tau} \langle \Psi(\bm{\theta}) | \hat{a}_{p\sigma}^{\dagger} \hat{a}_{q\tau}^{\dagger} \hat{a}_{s\tau} \hat{a}_{r\sigma} | \Psi(\bm{\theta}) \rangle.
\end{equation}
Because the quantum state remains frozen during the subsequent OO sweep, these RDMs are treated as constants.
The OO sweep then proceeds entirely on a classical computer. The orbital pairs are sequentially selected from the occupied-virtual subspace (i.e., occupied index \( i \in \{1, \dots, N_{\text{occ}}\} \) and virtual index \( a \in \{N_{\text{occ}}+1, \dots, N_{\text{orb}}\} \)). The pairs are visited {\color{black}in a fixed lexicographical order with respect to the indices \((i, a)\)}. For each target pair \( \{i, a\} \), the energy is evaluated at nine discrete Givens rotation angles \( \kappa_j = \frac{2\pi}{9} j \) (for \( j=0 \dots 8 \)) using the static RDMs and the locally rotated Hamiltonian integrals:
\begin{equation}
    \label{eq:energy_sweep}
    E_{ia}(\kappa_j) = \sum_{pq} \tilde{h}_{pq}(\kappa_j) \gamma_{pq} + \frac{1}{2} \sum_{pqrs} \tilde{h}_{pqrs}(\kappa_j) \Gamma_{pqrs},
\end{equation}
where \( \tilde{h}_{pq}(\kappa_j) \) and \( \tilde{h}_{pqrs}(\kappa_j) \) denote the one- and two-electron integrals temporarily updated by the local rotation angle \( \kappa_j \).
The Fourier coefficients \( A_0 \), \( A_n \), and \( B_n \) in Eq.~\eqref{eq:oo_fourier_expansion} are obtained from the sampled values \( E_{ia}(\kappa_j) \) by an inverse discrete Fourier transform.
Using the algebraic root-finding method described in Section~\ref{sec:aberth}, the global minimum \( \kappa_{ia}^{\text{opt}} \) of the reconstructed one-dimensional energy profile is determined for the target pair. The one- and two-electron integrals are immediately updated in-place, and the procedure advances to the next pair.

Once the classical OO sweep is completed, the updated integrals define a new transformed Hamiltonian. The algorithm then switches back to the quantum backend to optimize the amplitude parameters \( \bm{\theta} \). This is achieved using a similar sequential gradient-free strategy~\cite{nakanishi2020}; each parameter \( \theta_p \) is updated one-by-one in a fixed order by evaluating the energy at five distinct parameter values \( \theta_j = \frac{2\pi}{5} j \) (for \( j=0 \dots 4 \)) on the quantum circuit and identifying the minimum of the reconstructed one-dimensional energy profile. The optimized amplitudes are retained as the initial values for the subsequent macro-iteration.
This alternation between classical integral transformations and quantum ansatz optimizations can be repeated for a fixed number of macro-iterations or until an energy convergence criterion is met. The specific termination condition used in our numerical experiments is detailed in Section~\ref{sec:results}.
\section{Results \& Discussion}
\label{sec:results}
To evaluate the performance of the SOO-UCCD algorithm, we compute the potential energy surface (PES) of the lithium hydride (LiH) molecule (4 electrons, 6 spatial orbitals, STO-3G basis set) using exact state-vector simulation. All quantum circuit evaluations are performed classically without shot noise.
Electronic integrals and full configuration interaction (FCI) reference energies are obtained from PySCF~\cite{sun2015,sun2018,sun2020}.
The macro-iterations are repeated until the total energy change between consecutive iterations falls below a convergence threshold of \( 10^{-8} \) Hartree, or until a maximum limit of 100 macro-iterations is reached. We compare UCCD(seq), UCCSD(seq), and OO(seq)-UCCD(seq), where the labels in parentheses denote the optimizer used for the corresponding UCC amplitudes or orbital rotations. Here, ``seq'' denotes the sequential gradient-free update based on the reconstructed one-dimensional energy profile.
\begin{figure}[htbp]
    \centering
    \includegraphics[width=0.8\linewidth]{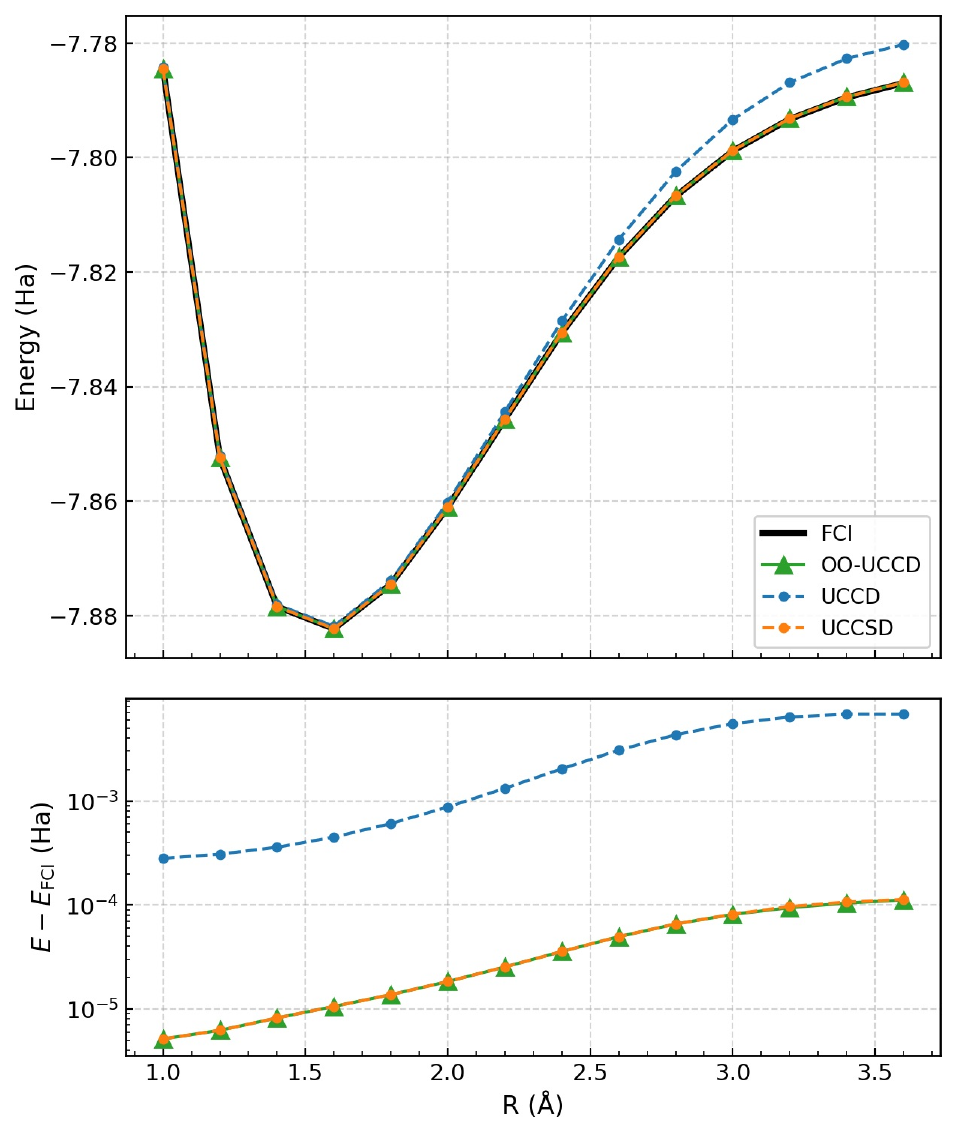}
    \caption{Potential energy surface of LiH (STO-3G). The top panel shows the total energy across various bond distances \( R \). The bottom panel shows the absolute energy error relative to the exact FCI energy. The PES curves use sequential optimization for the UCC amplitudes and, for OO-UCCD, for the orbital rotations.}
    \label{fig:lih_pes}
\end{figure}

Figure~\ref{fig:lih_pes} illustrates the calculated PES and the corresponding errors relative to the exact FCI solution. As the bond stretches (\( R > 2.0 \)~\AA), UCCD(seq), which lacks single excitations, exhibits significant deviations from the exact FCI energy, reaching errors on the order of \( 10^{-2} \) Ha. In contrast, UCCSD(seq) incorporates single excitations and accurately captures the correlation energy throughout the bond-stretching range considered.
OO(seq)-UCCD(seq) closely reproduces the accuracy of UCCSD(seq) and remains well within the conventional chemical-accuracy threshold, with errors on the order of \(10^{-5}\)--\(10^{-4}\) Ha across the curve.
This result is consistent with the role of orbital optimization in the OO-UCC framework~\cite{mizukami2020}: orbital rotations can account for much of the variational flexibility otherwise provided by single-excitation operators \( \hat{T}_1 \), allowing the shallower UCCD-only quantum circuit to achieve UCCSD-level accuracy in this system.
\begin{figure}[htbp]
    \centering
    \includegraphics[width=0.8\linewidth]{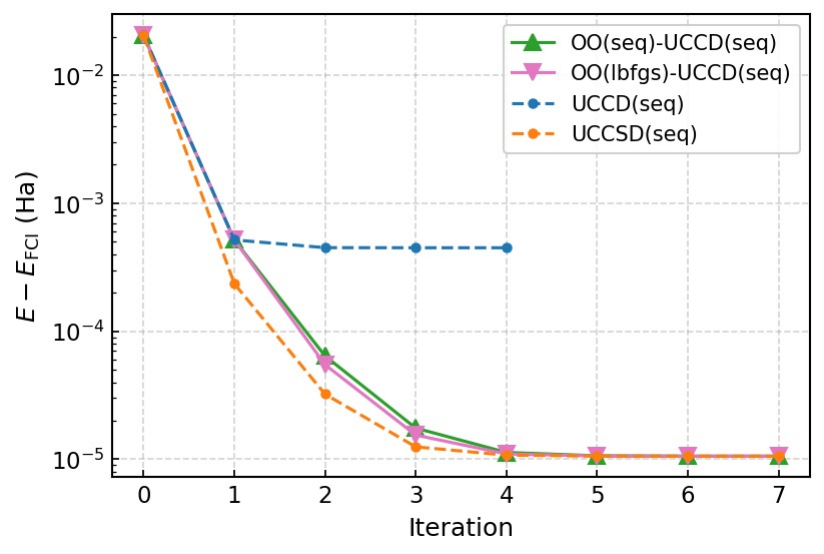}
    \caption{Macro-iteration convergence of the total energy for LiH near equilibrium geometry (\( R = 1.60 \)~\AA). The error is measured relative to the FCI ground state.}
    \label{fig:lih_convergence}
\end{figure}

For the convergence analysis, we also include OO(lbfgs)-UCCD(seq), in which the orbital rotations are optimized by the L-BFGS method from the \texttt{argmin} Rust optimization library~\cite{Kroboth2025}. Gradients are computed using central finite differences with step size \( \varepsilon = 10^{-5} \), which is appropriate for noise-free state-vector simulation.
Figure~\ref{fig:lih_convergence} details the convergence behavior of the macro-iterations near an equilibrium geometry (\( R = 1.60 \)~\AA). While UCCD(seq) quickly converges to a suboptimal local minimum, both OO(seq)-UCCD(seq) and OO(lbfgs)-UCCD(seq) converge to the UCCSD(seq) energy limit. The gradient-free sequential orbital sweep in OO(seq)-UCCD(seq) performs comparably to the L-BFGS orbital optimization in OO(lbfgs)-UCCD(seq), requiring 7 macro-iterations to reach convergence.
By achieving comparable accuracy without explicitly evaluating orbital gradients or Hessians, SOO-UCCD provides a gradient-free alternative for quantum--classical hybrid calculations.

These results highlight a practical feature of the SOO-UCCD algorithm relative to gradient-based orbital optimization such as the Newton--Raphson OO-VQE formulation~\cite{mizukami2020}. The sequential sweep is gradient-free and hyperparameter-free, yet shows macro-iteration convergence comparable to the L-BFGS orbital optimization in the present LiH benchmark.
Furthermore, it provides direct parameter updates from reconstructed one-dimensional energy profiles. From a hardware implementation perspective, shifting part of the variational flexibility associated with single excitations to classical orbital transformations allows the quantum backend to use a double-excitation-only ansatz without explicit single-excitation gates.
\section{Conclusion}
\label{sec:conclusion}
In this work, we have introduced SOO-UCCD, a fully sequential, gradient-free approach to orbital-optimized unitary coupled-cluster theory. By exploiting the analytical property that the energy expectation value is a fourth-degree trigonometric polynomial with respect to any single spatial orbital rotation, we demonstrated that the optimal rotation angle can be identified algebraically from nine energy evaluations. The proposed method is hyperparameter-free, performs a direct global search within each one-dimensional subspace, and avoids explicit orbital-gradient and Hessian evaluations.
Our numerical simulations of LiH show that SOO-UCCD achieves UCCSD-level accuracy, well within the conventional chemical-accuracy threshold, while using a double-excitation-only quantum ansatz. The sequential orbital optimization also shows macro-iteration convergence comparable to L-BFGS orbital optimization in the present benchmark.

By combining gradient-free sequential optimization with classical orbital transformations and a double-excitation-only quantum ansatz, SOO-UCCD provides a practical approach to orbital-optimized VQE for molecular electronic-structure calculations.

\section*{Acknowledgments}
This research was supported in part by Grants-in-Aid for Scientific Research
(Grant Nos. JP21K18903, JP22H05025, and JP25K01688) from the Ministry
of Education, Culture, Sports, Science and Technology (MEXT) of Japan.
This research was also partially supported by MEXT Quantum Leap
Flagship Program (MEXT Q-LEAP) Grant Number JPMXS0118067246
and JST COI-NEXT Grant Number JPMJPF2221. This work was also partly supported by IBM-UTokyo lab and RIKEN TRIP initiative.
OpenAI ChatGPT (GPT-5.6 Sol) was used during manuscript preparation
to assist with language editing. All AI-assisted suggestions were critically reviewed and independently
verified by the authors, who take full responsibility for the content of the manuscript.

\section*{Data Availability}
The data and source code supporting the findings of this article are
not publicly available at the time of publication because the source
code is under active development. They are available from the
corresponding author upon reasonable request.

\end{document}